%% file: main_draft_03WIP.tex
\documentclass[aps,prx,amsmath,amssymb,twocolumn,superscriptaddress,notitlepage,reprint]{revtex4-2}

\usepackage[utf8]{inputenc}
\usepackage[T1]{fontenc}

\usepackage{pgfplots}
\DeclareUnicodeCharacter{2212}{−}
\usepgfplotslibrary{groupplots,dateplot,fillbetween}
\usetikzlibrary{patterns,shapes.arrows}
\pgfplotsset{compat=newest}
\tikzset{
    semithick/.style={line width=0.8pt},
}

\usepackage{scalerel,stackengine} 
\usepackage{graphicx}
\usepackage{dsfont}
\usepackage{color}
\usepackage{epsfig}
\usepackage{bm}
\usepackage{times}
\usepackage{bbold}
\usepackage{bbm}
\usepackage{subfigure}
\usepackage[colorlinks,citecolor=blue,linkcolor=blue,urlcolor=blue]{hyperref}
\usepackage{scalerel,stackengine}
\usepackage{mathtools}
\usepackage{siunitx}
\usepackage{booktabs}
\usepackage{amsmath}
\usepackage{cleveref}

\usepackage{etoolbox} 
\usepackage{footmisc} 

\newcommand{\equalcontrib}{\thanks{These authors contributed equally. Email: nived.johny@aei.mpg.de}}
\newcommand{\m}{\mathrm{m}}
\renewcommand{\c}{\mathrm{c}}
\renewcommand{\a}{\mathrm{a}}

\newcommand{\om}{\mathrm{om}}

\newcommand{\BS}{\mathrm{\scriptscriptstyle{BS}}}
\newcommand{\DC}{\mathrm{\scriptscriptstyle{DC}}}

\newcommand{\rp}{\mathrm{\scriptscriptstyle{rp}}}

\definecolor{Black}{rgb}{0,0,0}

\begin{document}

\title{Realization of an all-optical effective negative-mass oscillator for coherent quantum noise cancellation}

\author{Nived Johny}
\equalcontrib
\affiliation{Max Planck Institute for Gravitational Physics (Albert Einstein Institute), D-30167 Hannover, Germany}
\affiliation{Leibniz Universit\"{a}t Hannover, D-30167 Hannover, Germany}

\author{Jonas Junker}
\equalcontrib
\affiliation{Max Planck Institute for Gravitational Physics (Albert Einstein Institute), D-30167 Hannover, Germany}
\affiliation{Leibniz Universit\"{a}t Hannover, D-30167 Hannover, Germany}
  \affiliation{OzGrav, Centre for Gravitational Astrophysics, Research School of Physics \& Research School of Astronomy and Astrophysics, Australian National University, Australian Capital Territory, Australia}
\affiliation{Center for Macroscopic Quantum States (bigQ), Department of Physics, Technical University of Denmark, 2800 Kongens Lyngby, Denmark}

\author{Bernd Schulte}
\affiliation{Max Planck Institute for Gravitational Physics (Albert Einstein Institute), D-30167 Hannover, Germany}
\affiliation{Leibniz Universit\"{a}t Hannover, D-30167 Hannover, Germany}

\author{Dennis Wilken}
\affiliation{Max Planck Institute for Gravitational Physics (Albert Einstein Institute), D-30167 Hannover, Germany}
\affiliation{Leibniz Universit\"{a}t Hannover, D-30167 Hannover, Germany}

\author{Klemens Hammerer}
\affiliation{Institut f\"{u}r Theoretische Physik, Universit\"{a}t Innsbruck, 6020 Innsbruck, Austria
}
\affiliation{Institute for Quantum Optics and Quantum Information, Austrian Academy of Sciences, 6020 Innsbruck, Austria}
\affiliation{Institute for Theoretical Physics, Leibniz Universit\"{a}t Hannover, Appelstra\ss{}e
  2, 30167 Hannover, Germany}

\author{Mich\`{e}le Heurs}
\affiliation{Max Planck Institute for Gravitational Physics (Albert Einstein Institute), D-30167 Hannover, Germany}
\affiliation{Leibniz Universit\"{a}t Hannover, D-30167 Hannover, Germany}
\date{\today}

\begin{abstract}

We report the realization of an all-optical, tabletop effective-negative-mass oscillator (ENMO) scheme capable of canceling quantum noise when cascaded with an opto-mechanical sensor susceptible to (quantum) radiation pressure noise. Our coherent quantum noise cancellation (CQNC) scheme offers a broadband cancellation capability with a tunable, wavelength-flexible, and compact system. This is achieved through the implementation of an optical equivalent of an opto-mechanical interaction, facilitated by a down-conversion and a beam-splitting process \cite{tsangCoherentQuantumnoiseCancellation2010}. The intricate nature of the system and its multiple interacting components made characterizing the interdependent parameters with conventional methods ineffective, leading to the development of an \textit{in‑situ} characterization scheme. The obtained parameters meet the targets for CQNC set in previous studies \cite{Wimmer2014}. With our current realization, we project a broadband quantum noise reduction of \qty{3.6}{dB}, corresponding to a \qty{77}{\percent} reduction in quantum back-action noise at the optimal frequency of maximum reduction, indicating the readiness of the ENMO for application. We discuss the prospects for new applications in quantum information and communication using the same platform.
\end{abstract}

\maketitle
\section{Introduction}
    An opto-mechanical interaction occurs when an optical light field reflects on a mechanical oscillator, a phenomenon that spans diverse scales \cite{aspelmeyerCavityOptomechanics2014}, from atoms in a light field \cite{ritschColdAtomsCavitygenerated2013,tebbenjohannsQuantumControlNanoparticle2021,monteiroForceAccelerationSensing2020}  to kilogram-scale mirrors like those used in LIGO and Virgo\cite{capoteAdvancedLIGODetector2025,ganapathyBroadbandQuantumEnhancement2023c,thevirgocollaborationQuantumBackactionKgScale2020,abacSearchContinuousGravitational2025}. These interactions have been employed in a wide range of applications, including fundamental science research \cite{chenMacroscopicQuantumMechanics2013,safavi-naeiniObservationQuantumMotion2012,murchObservationQuantummeasurementBackaction2008}, inertial sensors \cite{yver-leducReachingQuantumNoise2003}, gravitational wave detection \cite{braginskyInfluenceSmallDisplacement1964,braginskiiPonderomotiveEffectsElectromagnetic1967,cavesQuantumMechanicalRadiationPressureFluctuations1980,braginskyOpticalBarsGravitational1997,schnabelQuantumMetrologyGravitational2010,mcclellandAdvancedInterferometryQuantum2011,danilishinQuantumMeasurementTheory2012c}, and even for quantum computing \cite{reschBenchmarkingQuantumComputers2021} and quantum information processing \cite{barnumInformationTransmissionNoisy1998}. However, as the sensitivities of these systems have improved due to recent advances, they have become increasingly limited by quantum noise \cite{cavesQuantummechanicalNoiseInterferometer1981,cavesQuantumMechanicalRadiationPressureFluctuations1980}. To further improve the sensitivity of these systems, it is essential to address the optical power-dependent trade-off between quantum back-action (QBA) noise at lower frequencies and shot noise (SN) at higher frequencies. The limit caused by this trade-off is known as the standard quantum limit (SQL) \cite{cavesQuantummechanicalNoiseInterferometer1981,braginskyQuantumMeasurement1992,schnabelQuantumMetrologyGravitational2010}.

    Several schemes have been developed to achieve sub-SQL sensitivities. While frequency-independent squeezing has been proven effective for narrowband sensitivity enhancement \cite{abadieGravitationalWaveObservatory2011,loughFirstDemonstrationDB2021,aasiEnhancedSensitivityLIGO2013}, schemes like variational readout \cite{kimbleConversionConventionalGravitationalwave2002b,kampelImprovingBroadbandDisplacement2017} or frequency-dependent squeezing using filter cavities \cite{junkerFrequencyDependentSqueezingDetuned2022c,mccullerFrequencyDependentSqueezingAdvanced2020,ligoo4detectorcollaborationBroadbandQuantumEnhancement2023,virgocollaborationFrequencyDependentSqueezedVacuum2023} have been proven for broadband sub-SQL sensitivity. However, these approaches often require long filter cavities, which are impractical and costly to implement in a general quantum sensor, such as a small inertial sensor.



    Coherent quantum noise cancellation (CQNC) is an alternative scheme that enables broadband sub-SQL performance using a tabletop setup, and can be adapted for different quantum sensors \cite{tsangCoherentQuantumnoiseCancellation2010,Wimmer2014}. CQNC employs coherent feedforward control by introducing an anti-noise process to destructively interfere with the back-action noise. The opto-mechanical back-action process can be described as a down-conversion and beam-splitting process, which couples the optical field with the phononic excitation of the mechanical oscillator. The CQNC scheme employs an effective negative mass oscillator which mimics this process, but with an inverted sign. By cascading the ENMO and the opto-mechanical system (OMS), the back-action noise is reduced. While the initial proposal suggested an integration of positive and negative mass oscillators into a single system \cite{tsangCoherentQuantumnoiseCancellation2010}, later studies have shown that cascading both systems is possible and favorable \cite{schweerAllopticalCoherentQuantumnoise2022}. This cascadability allows the ENMO to be a separate system from the OMS, providing flexibility in the choice and tuning of its operation parameters.

     A cascaded CQNC implementation using atomic ensembles has been demonstrated successfully \cite{mollerQuantumBackActionevadingMeasurement2017}. The wavelength of operation is limited to the atomic transition wavelengths, which are often different from the wavelengths of interest for quantum sensors. This limitation necessitates an additional step of two-color Einstein–Podolsky–Rosen
    (EPR) entanglement \cite{brasilTwocolourHighpurityEinsteinPodolskyRosen2022a}. Further quantum‑non‑demolition (QND) measurements employing a cascaded reference‑frame architecture have been realized~\cite{woolleyTwomodeBackactionevadingMeasurements2013d,caniardObservationBackActionNoise2007d,mercierdelepinayQuantumMechanicsFree2021,hertzbergBackactionevadingMeasurementsNanomechanical2010a,ockeloen-korppiQuantumBackactionEvading2016d,hammererQuantumInterfaceLight2010,clerkBackactionEvasionSqueezing2008b,mollerQuantumBackActionevadingMeasurement2017}

    Here, we present an all-optical tabletop realization of an effective negative mass oscillator (ENMO) at a commonly used wavelength of \qty{1064}{nm}. The system parameters are precisely tunable, allowing for matching with a particular OMS. The Hamiltonian of an opto-mechanical interaction between an optical and a mechanical mode is recreated in our ENMO between two optical modes of two optical resonators. The coupling between these two optical modes is achieved through an optical 2-mode squeezing process and an independent beam-splitting process, enabling flexibility in the choice of wavelength and adaptability to different OMS configurations, including non-resonant operation. To overcome the challenge of operating a stable coupled optical resonator \cite{grafLengthSensingControl2013,steinmeyerSubsystemsAllopticalCoherent2019}, we employed a polarization-coupled optical resonator topology with a type II parametric down-conversion. 
    
    Due to the characterization challenge posed by the interdependent nature of system parameters, we developed a novel characterization method, which enabled the accurate determination of these parameters. Our new method enabled us to achieve the parameter values estimated in a previous publication \cite{Wimmer2014}, making the ENMO ready for cascading with an OMS (for instance, a membrane in the middle resonator with a Si$ _3$N$_4$ mechanical oscillator). From our measurements, we project that the ENMO could achieve a broadband cancellation of quantum back-action (QBA) noise of up to \qty{3.6}{dB} corresponding to a \qty{77}{\percent} reduction in quantum back-action noise at the frequency of maximum reduction when combined with the OMS. Furthermore, improvement in loss mitigation using state-of-the-art optical components shows great potential, enabling up to \qty{11.9}{dB} corresponding to an \qty{81}{\percent } QBA cancellation. Our realization of an ENMO will be further studied and modified for applications in quantum information, such as optical quantum memory and a single-photon source \cite{yoshikawaContinuousvariableQuantumOptical2017,yoshikawaCreationStorageOndemand2014}.

     \begin{figure}[]
        \centering
        \includegraphics[width=\linewidth]{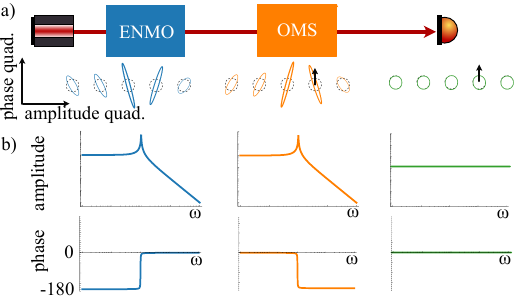 }
        \caption{ a) The cascaded CQNC system with the ENMO in series to an OMS and its corresponding quantum noise ellipses at different frequencies are shown for each system. The tailored frequency-dependent inversely squeezed state (blue) destructively interferes with the ponderomotively squeezed state (orange), resulting in vacuum states (green). The scheme reveals the force signal (black arrow), which was masked by the additional phase quadrature noise due to ponderomotive squeezing. b) shows the schematic plots for the ideal susceptibilities of the oscillators.} 
        \label{fig: Intro}
    \end{figure}

 \section{Theoretical background}

        Quantum back-action noise in an optomechanical system appears as an excess of phase‑quadrature fluctuations in the light reflected from a mechanical oscillator. These unwanted phase fluctuations arise because the fluctuating radiation‑pressure force exerted by the incident light displaces the oscillator. Since the radiation‑pressure force is proportional to the optical power, it is linked to the amplitude quadrature of the incoming field. Consequently, the phase quadrature of the reflected light becomes correlated with the amplitude quadrature, generating a ponderomotive squeezing of the reflected field \cite{braginskyInfluenceSmallDisplacement1964,braginskiiPonderomotiveEffectsElectromagnetic1967a,corbittSqueezedstateSourceUsing2006a,danilishinAdvancedQuantumTechniques2019}. In this process, the anti-squeezing noise projection  masks the desired signal in the phase quadrature. The frequency dependence of the effect is governed by the mechanical displacement susceptibility of the oscillator to the applied radiation‑pressure force.


    To perfectly cancel ponderomotive squeezing, the ENMO must provide an opposite, frequency‑dependent interaction that generates an “inversely squeezed” state in the light entering the OMS (Fig.~\ref{fig: Intro}a)). This inverse squeezing undoes the optomechanical squeezing, restoring a vacuum‑like output and revealing the signal~\cite{kimbleConversionConventionalGravitationalwave2002b}. As shown in Fig.~\ref{fig: Intro}b), the mechanical susceptibilities of the two oscillators must be matched in magnitude but opposite in phase. The theoretical basis for this cancellation scheme is presented by Tsang and Caves~\cite{tsangCoherentQuantumnoiseCancellation2010}, while Refs.~\cite{Wimmer2014, schweerAllopticalCoherentQuantumnoise2022} discuss its benefits, limitations, and realistic case studies.

    

    Here, we summarize in a three-step process the theoretical background for our implementation of an all-optical ENMO where a high-finesse cavity, the ancilla cavity, is the analog of the mechanical oscillator. First, we mimic the radiation pressure interaction by reproducing its Hamiltonian in the optical domain. Second, we will match the susceptibilities of the mechanical oscillator in the OMS and the corresponding optical resonator of the ENMO with opposite phase. Third, we will couple these oscillators in a similar way by matching the readout optical cavities and the coupling strengths.
    
\begin{figure}[h]
            \centering
            \includegraphics[width=\linewidth]{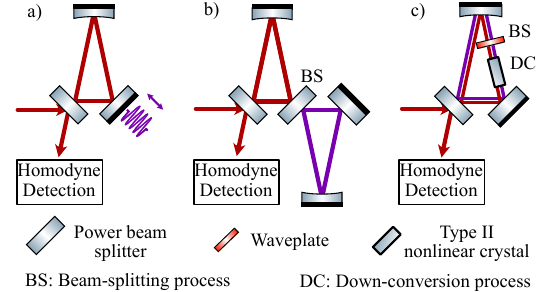}
            \caption{a) Representation of a typical opto-mechanical cavity. Here, the optical resonator (meter cavity) field shown in red interacts with the mechanical oscillator mirror shown in purple. The output field from the opto-mechanical cavity has QBA noise imprinted (ponderomotive squeezing), which is detected using a homodyne detection scheme. b) Here, the opto-mechanical oscillator is replaced by a high-finesse optical resonator (ancilla cavity) shown in purple. The power beam-splitter couples the two oscillators through the beam-splitting process. The down-conversion process is not present here.  c) In this realization, the high-finesse ancilla cavity is an orthogonal polarization mode instead of a spatially separated cavity. The opto-mechanical interaction is then mimicked by an optical-optical interaction by using a type II down-conversion crystal (down-conversion process) and a waveplate (beam-splitter process), and careful selection of the oscillator configuration.}
            \label{fig: Theory}
        \end{figure}

    The radiation pressure interaction Hamiltonian of an opto-mechanical cavity as shown in Fig.~\ref{fig: Theory}a) is \cite{schweerAllopticalCoherentQuantumnoise2022,berndwolfgangschulteCharacterisationIntegrationOptomechanical2023}    
    \begin{align}\label{Hrp}
    \begin{split}
      H_\rp & =  \frac{g_\om}{2}   \left[(b c_\om + b^\dagger c^\dagger_\om) + (b c^\dagger_\om + b^\dagger c_\om)\right] ,
    \end{split}
    \end{align}
     where $g_\om$ is the opto-mechanical coupling strength, $c_\om$ ($c_\om^\dagger$) and $b_\om$ ($b_\om^\dagger$)  are the annihilation (creation) operators of the optical cavity mode and the mechanical phonon mode, respectively. The first term of the Hamiltonian denotes a two-mode squeezing process, and the second term denotes beam-splitting processes between the optical mode $c$ and the mechanical mode $b$. To emulate the same interaction all-optically in a photon-photon mode, one can literally use a beamsplitter and a down-conversion process. As shown in Fig.~\ref{fig: Theory} b), the mechanical oscillator is replaced with an optical oscillator with the power beam-splitter achieving the beam-splitting interaction. The meter cavity $c$ (depicted in red) and the ancilla cavity $a$ (depicted in purple). The 2-mode squeezing process is not represented here.

    However, for ease of experimental realization and also to introduce the 2-mode squeezing process, a polarization-coupled cavity scheme is chosen instead of a power-coupled cavity as shown in Fig.~\ref{fig: Theory} c). In this case, the waveplate acts as a beam-splitter coupling power between the two orthogonal polarization modes (see  Appendix.~\ref{waveplateBS}). A type  II nonlinear crystal is used to realize the 2-mode squeezing process between the two polarization modes. 
       
     The susceptibilities of the mechanical oscillator $\chi_\m$ and the ancilla cavity $\chi_\a$ dictate the frequencies at which the above-mentioned interaction occurs \cite{Wimmer2014}. Therefore, they are equalized, but with the opposite phase $\chi_\m = -\chi_\a $. This gives the following conditions \cite{schweerAllopticalCoherentQuantumnoise2022}:

       \begin{subequations}
     \begin{align}
        \Delta_\a &=- \omega_\m , \label{Eq:susc1}\\ 
        \kappa_\a  &= \gamma_\m \label{Eq:susc2}  \\ 
         |\Delta_\a| &\gg \kappa_\a \label{Eq:susc3}
     \end{align}\label{Eq:susc}
    \end{subequations}

    where $\Delta_\a$ and $\kappa_\a$ are the detuning and linewidth of the ancilla cavity, respectively, and $\omega_m$ and $\gamma_m$ are the resonance frequency and linewidth of the mechanical oscillator, respectively.
    
    To ensure equal measurement strength of the ENMO and the OMS, the susceptibilities $\chi_c$  and $\chi_\om$ of the respective readout cavities must be matched. This gives the condition,

    \begin{equation}       
        \kappa_\c = \kappa_\om, \label{eq:meter_cavities}
    \end{equation} 
    where $\kappa_\om$ is the linewidth of the optomechanical meter cavity and $\kappa_\c$ is the linewidth of the ENMO meter cavity. 
    The coupling of the meter cavities to these oscillators are matched with the following condition with their single-photon  coupling strengths   
    \begin{subequations}
    \begin{align}        
        g_\a &=g\BS+ g_\DC = g_\om, \label{Eq:coupl1} \\
        g\BS &= g_\DC, \label{Eq:coupl2}
    \end{align}\label{Eq:coup}%
    \end{subequations}
    where $g_\a$ is the coupling strength of the ENMO, $g_\BS$ is the beam-splitting coupling strength, and $g\DC$ is the down-conversion coupling strength.
    
    With these matched, the interaction of the opto-mechanical cavity with the phononic mode is similar to the interaction of the meter cavity with the photonic mode of the ancilla cavity. The two systems are entangled.

    The inversely‑squeezed state is generated when the pump field undergoes a two‑mode squeezing interaction, producing a pair of orthogonally polarized, quantum‑correlated fields. Within the ENMO the photons in the ancilla cavity are coherently exchanged with those in the meter cavity via a beam‑splitting interaction; for perfect cancellation the two‑mode‑squeezing rate must equal the photon‑exchange rate. Only photons exiting the meter cavity are available for detection. The relative phase between the two correlated photons determines the rotation of the squeezed quadrature; this phase acquires a frequency dependence because the ancilla cavity is detuned as in Eq.~\ref{Eq:susc1}. Likewise, the asymmetric overlap of the amplitude susceptibilities of the meter and ancilla cavities imparts a frequency‑dependent modification of the squeezing magnitude. Consequently, the combined frequency‑dependent squeezing factor and squeezing angle constitute the tailored inversely‑squeezed state required for back‑action evasion.

    
    The phase quadrature spectral density of the inversely   squeezed state in an ideal CQNC scheme is~\cite{schweerAllopticalCoherentQuantumnoise2022,berndwolfgangschulteCharacterisationIntegrationOptomechanical2023}    
    \begin{align}\label{Eq:variance}
    \begin{split}
    S(\omega)_{\mathrm{CQNC}} &= \frac{1}{2} 
    +\frac{G_\a^2}{2 } \left\lvert \chi_\m +\chi_\a\right\rvert^2 \\
    &\quad+\frac{G_\a \kappa_\a |\chi_\a|^2}{2 }\left(\frac{\omega^2 + \kappa_\a^2/4}{\Delta_\a^2}+1\right),
    \end{split}
    \end{align}    
    where $G_\a= g_\a^2 \kappa_\a \lvert \chi_\a(\omega) \rvert^2$ is the frequency-dependent measurement strength of the ENMO. The corresponding  OMS's measurement strength is $G_\om= g_\om^2 \kappa_\om \lvert \chi_\om(\omega) \rvert^2$. In an ideal CQNC scheme, with perfect matching, the second term vanishes because $\chi_\a =- \chi_\m$. The remaining terms are the shot noise of the incoming light field (first term) and the vacuum noise coupling due to the losses in the ancilla cavity (last term). These then determine the noise floor of the CQNC scheme. 

    The study of \cite{schweerAllopticalCoherentQuantumnoise2022} concludes that for the best performance, the EMNO should precede the quantum sensor. It studies the feasibility of realistic conditions and the effects of mismatches of parameters. It was shown that even with a realistic mismatch of the mechanical linewidth $\gamma_m$ with the ancilla cavity optical linewidth $\kappa_\a$ in Eq.~\eqref{Eq:susc2}, broadband sub-SQL performance is possible at off-resonance frequencies. Any mismatch between $g_\BS$ and $g_\DC$ drastically reduces the sensitivity at frequencies above the resonance. Therefore, precise control of $g_\BS$ and $g_\DC$ in Eq.~\eqref{Eq:coupl2} is important. Additionally, even for a parameter-mismatched system, a narrow-band enhancement of sensitivity other than at resonance is possible with a specific combination of $\kappa_c$, $ g_\BS$  and $g_\DC$~\cite{schweerAllopticalCoherentQuantumnoise2022}.

 \section{Experiment}

     \begin{figure}[!h]
            \centering
            \includegraphics[width=\linewidth]{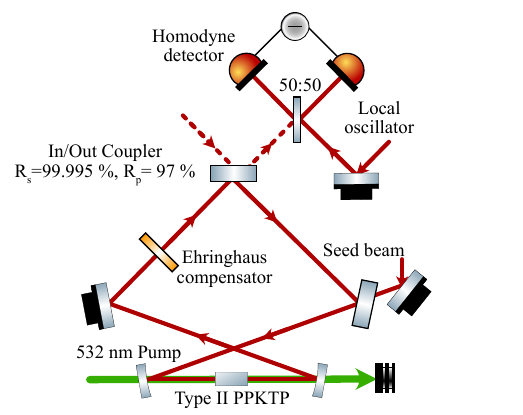}
            \caption{Schematic of the optical realization of the ENMO. It consists of a polarization-coupled cavity system with a type II PPKTP crystal. The cavity is \qty{1.52}{m} long and has an FSR of \qty{197.4}{MHz}. The input/output coupler has a reflectivity of \qty{97}{\percent} for p-polarization and reflectivity of \qty{99.995}{\percent} for s-polarization.  The cavity is pumped with a single-pass \qty{532}{nm} light field in s-polarization. The dashed lines represent the vacuum states in p-polarization that are squeezed by the setup. Homodyne detection with a local oscillator in p-polarization detects the inversely squeezed states in the p-polarized light.}
            \label{fig: Setup}
        \end{figure}

    Experimentally realizing the constraints given by Eqs.~\labelcref{Eq:susc,eq:meter_cavities,Eq:coup} is challenging. Most considered topologies \cite{steinmeyerSubsystemsAllopticalCoherent2019} failed for the following reasons:  1. Realization of the beam-splitting process necessitates a coupled-cavity system. Stable operation of the system is difficult owing to the coupling; in particular, locking the lengths of the sub‑cavities is challenging because of their interaction. 2. Each sub-cavity requires different line-widths to satisfy Eq.~\eqref{Eq:susc2} and Eq.~\eqref{eq:meter_cavities} while keeping the losses to a minimum. To avoid decoherence due to losses, the topology and the components have to be chosen carefully. 3. The coupling strengths $g_\BS$ and $g_\DC$ have to be precisely and independently tunable to satisfy Eq.~\eqref{Eq:coupl1} and Eq.~\eqref{Eq:coupl2}. Satisfying Eq.~\eqref{Eq:coupl1} is crucial to match the ENMO to the OMS and satisfying Eq.~\eqref{Eq:coupl2} for high-frequency QBA cancellation \cite{schweerAllopticalCoherentQuantumnoise2022}. Furthermore, easy modification of the cavity detuning $\Delta_\a$ is necessary to facilitate adaptation to a different mechanical oscillator resonance frequency (refer to Eq.~\eqref{Eq:susc1}).


   To tackle the above challenges, firstly, we chose a polarization-based coupled cavity system as shown in Fig. \ref{fig: Setup}, where the two orthogonal polarization modes (p-pol and s-pol) correspond to the two modes of the ENMO. P-pol mode is the meter cavity mode. S-pol mode is the ancilla cavity mode, which mimics a mechanical oscillator. Locking the length of the two sub-cavities is then simplified to locking the same physical mirrors by the Pound-Drever-Hall technique in reflection using a counter-propagating beam (not shown in Fig. \ref{fig: Setup}). Further, the relative detuning, and thus the relative cavity length, between the two polarizations could be tuned by changing the relative retardation in a birefringent bulk in the cavity. The birefringent nature of the type II nonlinear periodically poled
    potassium titanyl phosphate (PPKTP) crystal used for the  down-conversion process is exploited for this by changing its temperature  \cite{junkerFrequencyDependentSqueezingDetuned2022c,junkerReconstructingGaussianBipartite2022,junkerSqueezingNormalModeSplitting2025}. The crystal's polarization basis must be perfectly matched with that of the cavity to avoid the crystal undesirably introducing beamsplitter-like mode-coupling. This is important for the  precise control of the coupling strength $g_\BS$. Appendix~\ref {crystal0} contains a more detailed explanation and techniques developed for precise execution. The setup is enclosed in a box to reduce acoustic, thermal, and air fluctuations, thus allowing for sufficient stability.  

    Secondly, to achieve different line-widths $\kappa_\a$ and $\kappa_c$ for the two cavity modes  while keeping losses minimal, an input/output coupler with different reflectivities for the two polarizations was chosen, while all other mirrors are high-reflective (HR) mirrors for both polarizations. In this way, the dominant part of the p-pol cavity decay is detected, thereby increasing the escape efficiency while fulfilling the requirements set by Eqs.~\ref{Eq:susc2},\ref{eq:meter_cavities}. The s-pol cavity is essentially realized with  HR mirrors, with its dominant useful decay channel being the beam-splitting process, coupling it to the p-pol cavity. The different reflectivity for p-pol \qty{97}{\percent} and s-pol  \qty{99.995}{\percent} is achieved by using an HR coated mirror for \qtyrange{4}{8}{\degree} angle of incidence (AOI) in a \qty{45}{\degree} AOI configuration. This creates the unorthodox shape of the cavity as shown in Fig.~\ref{fig: Setup}. Furthermore, the crystal and the waveplate in the cavity are anti-reflective (AR) coated for \qty{1064}{nm} wavelength. Further improvement of our system performance could be achieved by using the latest technology of AR-coatings (<\qty{0.01}{\percent}~\cite{LaseroptikCatalog2025}) as it becomes available to us, because this is our dominant loss channel. The losses due to the HR mirrors are negligible.

    Lastly, the choice of a waveplate instead of an actual beam-splitter facilitates the tunability of $g_\BS$. In previous schemes, the beam-splitter was realized by a conventional power splitter and, accordingly, $g_\BS$ could not be tuned~\cite{Wimmer2014,tsangCoherentQuantumnoiseCancellation2010}. The beamsplitter coupling strength for a waveplate is given by $g_\BS = \sin(2\delta)\,  \theta\, \textit{FSR}$ \cite{steinmeyerSubsystemsAllopticalCoherent2019}, where $\theta$ is the relative phase shift between the fast and slow axis of the waveplate, $\delta$ is the angle of the waveplate optical axis to the s-pol field and \textit{FSR} is the free spectral range of the cavity.  In previous attempts of polarization-coupled cavities, precise control of $g_\BS$ was difficult due to the use of a typical half- or quarter-wave plate where $\theta =\pi,~\frac{\pi}{2}$ is large and fixed \cite{steinmeyerSubsystemsAllopticalCoherent2019}. Instead, in our case, an Ehringhaus compensator \cite{burriZurTheorieUnd1953} was used, where we have $\theta \approx 0$. Tilting the Ehringhaus compensator gives a degree of freedom to change $\theta$, thereby changing the sensitivity of $g_\BS$ tuning under rotation of $\delta$.

     The noise variance measurements were taken at the first \textit{FSR} due to excess power noise of our laser at baseband. We used a \qty{532}{nm} pump field in a single-pass configuration with a pump power of less than \qty{140}{mW}. The \qty{1064}{nm} coupled cavity has a p-pol linewidth measured to be $\kappa_c \approx 2\pi \times$ \qty{1}{MHz}, and the s-pol linewidth was inferred from the measurement fits to be $\kappa_a \approx 2\pi \times$ \qty{160}{kHz}. Via an HR mirror, a seed beam is coupled in to weakly displace the vacuum-squeezed states to a bright squeezed state. The p-pol squeezed state is detected using a balanced homodyne detection scheme with an \qty{11}{mW} p-pol local oscillator. 

     The AC part of the homodyne detection signal is sent to a spectrum analyzer to measure the noise variance, and the DC part of the signal is sent to an oscilloscope. The DC part depends on the interference between the seed field and the local oscillator field, from which the detection angle can be extracted and controlled. A full schematic of the system, including the locking schemes, is explained in Appendix~\ref{locks}.

 \section{Characterization}

          \begin{figure}[]
            \centering
            \includegraphics[width=\linewidth]{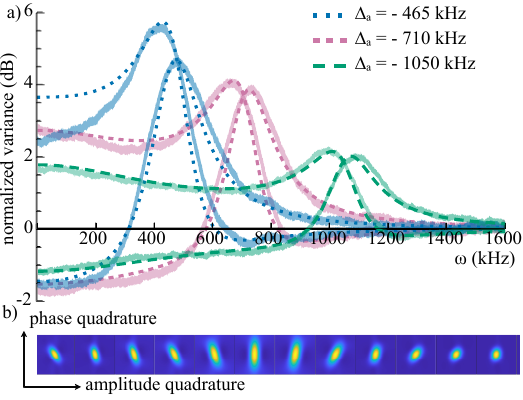}
                \caption{a) Measured variance (normalized to the shot noise, with dark noise subtracted) of the output field as a function of frequency for the ancilla cavity detuning $\Delta_\a$ of \qty{-465}{kHz} (light blue), \qty{-710}{kHz} (light red) and \qty{-1050}{kHz} (light green). For each detuning, the variance is shown for two approximately orthogonal detection-locked quadratures. The corresponding fits are shown as dashed lines in matching dark colors. The squeezing measurements correspond to an ENMO for a Si$ _3$N$_4$ membrane opto-mechanical system as compared in Table. \ref{tab: exp output} b) Tomographical reconstruction of the quantum squeezed ellipse over frequency for $\Delta_\a$ of \qty{-710}{kHz}. }

            \label{fig: Variance_fits}
        \end{figure}

  \begin{table}[!h]
            \centering
               
            \begin{tabular}{ l r | l r }\toprule
            ENMO   & Value/2$\pi$   & OMS  & Value/2$\pi$ \\ \midrule     
            $ \Delta_c $ & \qty{0}{kHz} $\pm$ \qty{3}{kHz}&$ \Delta'_\om$  & 0 MHz\\ 
            $ \Delta_\a $ &  \qty{-465}{kHz} $\pm$ \qty{3}{kHz}  & $\omega'_{m}$ & $\approx$\qty{417}{kHz}   \\
            &\qty{-710}{kHz}$\pm$ \qty{3}{kHz}&&$\approx$\qty{660}{kHz}\\
             &\qty{-1050}{kHz}$\pm$ \qty{3}{kHz}&&$\approx$\qty{1065}{kHz}\\
            $ \kappa_\a$  & \qty{160}{kHz}$\pm$ \qty{20}{kHz} &$ \gamma'_{m}$ & $\approx$\qty{1}{Hz} \\ 
            $ \kappa_c $ & \qty{980}{kHz} $\pm$ \qty{50}{kHz} &$\kappa'_{om}$ & $\approx$\qty{1}{MHz} \\ 
            $ g_{BS}$  & \qty{175}{kHz} $\pm$ \qty{5}{kHz}& $-$ & $-$\\ 
            $ g_{DC}$  & \qty{175}{kHz} $\pm$ \qty{5}{kHz} &  $-$ & $-$\\ 
            $ g_{BS}+g_{DC}$  & \qty{350}{kHz} $\pm$ \qty{10}{kHz}&$ g'_{om}$ & \qty{350}{kHz} \\ \bottomrule
            \end{tabular}
            
           \caption{Shows the values extracted from the fit corresponding to Fig \ref{fig: Variance_fits}. The parameters of the ENMO and its corresponding OMS equivalent are shown. This was chosen corresponding to a membrane in the middle opto-mechanical cavity with a Si$ _3$N$_4$ mechanical oscillator, which is available in our lab.} 
           \label{tab: exp output}
        \end{table}

      
     Methods to individually characterize $g_\BS$ and $g_\DC$ do exist for a simpler system \cite{steinmeyerSubsystemsAllopticalCoherent2019}. However, the system here with a coupled cavity of different linewidths ($\kappa_c $ and $\kappa_\a$)  and detuning ($\Delta_c$ and $\Delta_\a$) for the two optical modes makes those methods impractical to implement. Furthermore, setting the experiment in different configurations to measure each system parameter is time-consuming and inaccurate.  Therefore, we developed an \textit{in situ} measurement and fitting routine to quickly characterize the coupling strengths ($g_\BS$ and $g_\DC$), cavity line-widths ($\kappa_c $ and $\kappa_\a$), and detunings ($\Delta_c$ and $\Delta_\a$). This is helpful during applications when the ENMO is to be operated alongside a particular OMS.

    Fig.~\ref{fig: Variance_fits} a) shows measurement trace (light) pairs of 3 different detunings $\Delta_\a$ and their corresponding fits (dashed dark) of the variance over frequencies. Each color pair corresponds to locked detection angles, which are approximately \qty{90}{\degree} apart. The fit model is obtained by setting $\chi_m = 0$ in Eq.~\ref{Eq:variance} and implementing losses and rotation as in Appendix~\ref{Ap:covariance}. These measurements were taken with a Keysight N9020A spectrum analyzer (resolution bandwidth (RBW) \qty{100}{kHz}; video bandwidth (VBW) \qty{12}{Hz}; sweep time \qty{1.01}{s}; average 10). Because the measurements are taken at the first FSR, the measurement with the lowest $\Delta_\a$ measurement (blue) is much more prone to effects due to $\mathrm{FSR}_s \neq \mathrm{FSR}_p$, and thus deviates from the fits at low frequencies. For baseband detection, this effect does not occur. For the case of $\Delta_\a=$ \qty{-710}{kHz}, a quantum tomography reconstruction of the squeezed ellipse at different frequencies is also shown in Fig.~\ref{fig: Variance_fits} b) \cite{junkerFrequencyDependentSqueezingDetuned2022c}. The squeezing factor and angles change over frequency, thus creating a tailored inversely squeezed state that can coherently and destructively interfere with a ponderomotively squeezed state from an OMS.

    Table~\ref{tab: exp output} summarizes the results of the measurement fits. The ENMO was operated considering future operations alongside a membrane in the middle cavity with a Si$ _3$N$_4$ (silicon nitride) membrane, whose parameters are also shown. Here, we opted for a resonant OMS;  therefore, the p-pol cavity is locked on resonance. The measurements are taken for 3 different  s-pol cavity detunings $\Delta_\a$ where our  Si$ _3$N$_4$  membrane (\qty{500}{\micro \meter} or \qty{1000}{\micro \meter} side length) has its resonances. The value $\kappa_\a = 2\pi \times$ \qty{160}{kHz} is dominated by the losses in the cavity caused by the non-perfect anti-reflective coatings of the PPKTP crystal and the Ehringhaus compensator. The $g_\BS = 2\pi \times$ \qty{175}{kHz} was finely tuned to be equal to $g_\DC$ using a motorized waveplate rotation stage. This corresponds to an opto-mechanical system with $g_\om = 2\pi \times$  \qty{350}{kHz}. The values obtained are consistent with the targets set in the past study~\cite{Wimmer2014}. 

   The system's losses can be characterized by operating it as a single-mode squeezer. This is done by setting the waveplate to couple maximally between the s-pol and p-pol modes (i.e., in the analogy of a coupled cavity realization using power beam-splitter, the reflectivity of it is set to zero), thus effectively doubling the cavity length due to the high finesse of the s-pol cavity. In this configuration, the system could be considered as a simple single-mode squeezer to check for unaccounted losses \cite{vahlbruchDetection15DB2016}. 

 \section{Discussion and Outlook}

         \begin{figure}[!h]
            \centering
            \includegraphics[width=\linewidth]{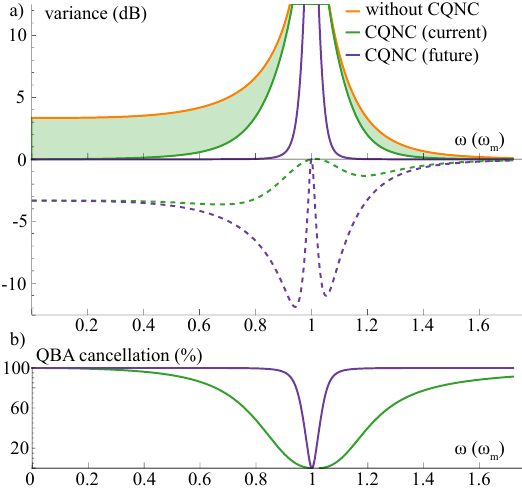}
            \caption{a) Simulated variance of the noise  in the phase quadrature due to ponderomotive squeezing (orange line) and the potential broadband cancellation with the current ENMO (green line), normalized to shot noise. These are shown for the OMS and ENMO parameters given in Tab. \ref{tab: exp output}. The green-shaded area shows the reduction in noise. The projected noise with a potential future ENMO with improvement of $\kappa_\a =$\qty{10}{kHz} is shown in purple.  The corresponding dashed lines show the canceled noise. b) shows the percentage of QBA noise that could be canceled with the current and future projected ENMO realization.}
            \label{fig: simulated_ellipses}
        \end{figure}

    Combining the measured ENMO with a simulated OMS, the noise reduction due to CQNC can be projected as in Fig.~\ref{fig: simulated_ellipses}. Fig.~\ref{fig: simulated_ellipses}a) shows in orange the phase quadrature noise increase over frequency due to quantum back action noise for an opto-mechanical system with the parameters shown in Table~\ref{tab: exp output}. The green curve represents the reduced noise due to CQNC dictated by Eq.~\ref{Eq:variance} with the currently achieved ENMO parameters. The ENMO suppresses noise for all off-resonance frequencies even with a mismatch of $\gamma_\m$ and $\kappa_\a$. The corresponding force spectral density could also be calculated as in \cite{schweerAllopticalCoherentQuantumnoise2022}. The purple curve represents the noise that could potentially be achieved with a future improved ENMO with reduced ancilla cavity linewidth $\kappa_\a = 2\pi \times$ \qty{10}{kHz}. The dotted lines represent the amount of noise that is reduced by the current (and future) ENMO over frequency. Fig.~\ref{fig: simulated_ellipses}b) shows the percentage of QBA noise canceled over frequency. With the current ENMO, a maximum noise reduction of \qty{3.6}{dB} corresponding to \qty{77}{\percent} could be achieved at frequencies approximately $0.67\times\omega_m$. With $\kappa_\a = 2\pi \times $\qty{10}{kHz}, a maximum noise reduction of \qty{11.9}{dB} corresponding to \qty{81}{\percent} could be achieved at frequencies approximately $0.94 \times \omega_m$.
    
     The easy tunability of the parameters, such as $\Delta_\a$, $g_\BS $ and $g_\DC$, makes the system easy to adapt for different OMS.  $g_\BS$ can be arbitrarily increased up to the cavity's FSR. $\Delta_\a$ could be increased desirably; however, for a constant $g_\DC$ and $g_\BS$, the optical parametric oscillation threshold increases with higher detuning, increasing the demand on pump power. Simply increasing the pump power for higher  $g_\DC$ leads to the optical parametric oscillation, setting a limit to the maximum achievable $g_\DC$. Contrarily, keeping the pump power constant instead, increasing the detuning decreases the squeezing and anti-squeezing values. With the current system, a tuning range of $0 < g_\DC < 2\pi \times$ \qty{175}{kHz} is possible. The  system could also be used in conjunction with a non-resonant OMS by accordingly detuning the p-pol cavity and changing the $ g_\DC/ g_\BS$ ratio. Furthermore, \cite{schweerAllopticalCoherentQuantumnoise2022} has also shown configurations where another narrowband sensitivity peak could be achieved further away from the resonance with a specific choice of mismatched parameters. 

    Now that the full ENMO system has been studied and proven to be suitable in this prototype setup, further improvements could be targeted in the future. $\kappa_\a$ could be decreased by increasing the cavity length. Quadrupling the cavity length to \qty{6}{m} could still be achieved as a tabletop experiment. Secondly, exploiting the birefringence of the PPKTP crystal to use as a waveplate, one can eliminate two AR coated surfaces, which are the highest contributors to loss. Furthermore, choosing the latest available AR coatings for the crystal and HR coating of the mirrors altogether, one can realistically achieve $\kappa_\a$ on the order of $2\pi \times $\qtyrange{5}{20}{kHz}. This corresponds to a photon lifetime of \qtyrange{0.05}{0.2}{ms} in the ancilla cavity.

    Our system is not only suitable as an ENMO for CQNC, but can also serve as a platform where a high-finesse oscillator with a high decoherence time is coupled to a low-finesse cavity by an easily tunable down conversion and beam-splitting process. Potential applications in quantum information could be a quantum memory or a single photon source \cite{yoshikawaContinuousvariableQuantumOptical2017,yoshikawaCreationStorageOndemand2014,lvovskyOpticalQuantumMemory2009,makinoSynchronizationOpticalPhotons2016,pittmanSinglePhotonsPseudodemand2002,tanabeDynamicReleaseTrapped2009,leiQuantumOpticalMemory2023}. Choosing between the down-conversion process and the beam-splitting process to store and retrieve information is emulated in an OMS using the detuning of the pump beam. Replacing our current slow motorized waveplate with a faster one (e.g., an electro-optic modulator) and a quick switch (e.g., an acousto-optic modulator) in the pump beam path will enable us to quickly switch between $ g_\DC$ and $ g_\BS$ \cite{wallucksQuantumMemoryTelecom2020,kristensenLonglivedEfficientOptomechanical2024}. Due to the optical nature of our platform with no mechanical oscillators, it does not suffer from thermal decoherence, which is a limiting factor in opto-mechanical memories, and therefore can be operated at room temperature. This makes it easier to set up and operate than other realizations \cite{wallucksQuantumMemoryTelecom2020b,kristensenLonglivedEfficientOptomechanical2024,choHighlyEfficientOptical2016,tanabeTrappingDelayingPhotons2007}. Other promising optical memories and single photon sources have already been shown in \cite{yoshikawaContinuousvariableQuantumOptical2017,yoshikawaCreationStorageOndemand2014,lvovskyOpticalQuantumMemory2009,makinoSynchronizationOpticalPhotons2016}. Future research on our platform for these types of applications could offer beneficial perspectives.

 \section{Conclusion}
    
    We have presented the first successful realisation of an all-optical ENMO capable of broadband reduction of QBA noise in an OMS. We have demonstrated the first experimental realization  that has overcome previously limiting challenges of coupled cavities and have realized precise tunability. The newly developed characterisation techniques provide an easy and quick way of determining the system parameters. We have achieved values anticipated in a previous publication \cite{Wimmer2014} and, thus, our ENMO is ready for integration with a matched OMS. The tunability of the system parameters makes it adaptable for different OMSs and their operating points. The noise reduction for a potential quantum sensor with the current ENMO system and a potential future ENMO system is shown. Finally, in addition to an ENMO ready to use for CQNC, we also offer to the community the possible use of this platform with slight modifications for quantum information and communications applications. 

\begin{acknowledgments}
This work was funded by the Deutsche Forschungsgemeinschaft Excellence
QuantumFrontiers (EXC 2123, Project ID 390837967). We would like to express our gratitude to Asst. Prof. Nenad Kralj for valuable discussions and insights that contributed to this work. We thank Dr. Mikhail Korobko and Dr. Fabio Bergamin for providing valuable feedback.
\end{acknowledgments}
  
 \begin{appendix}
    \section{beam-splitter interaction} \label{waveplateBS}

    \subsection{Waveplate as a beamsplitter} 
    The coupling interaction of two input modes to become two output modes is described mathematically using a coupling matrix. The coupling matrix of a beam-splitter can be represented as
       \begin{equation}
    M_{\text{beamsplitter}}=
    \begin{pmatrix}
t & r \\
r' & t'
\end{pmatrix} 
  \end{equation}
  where, they follow the following relations
     \begin{subequations}
    \begin{align}        
        |r| =|r'|, |t| &=|t'|, \label{eq:BS1} \\
        |r|^2 +|t|^2&=1, \label{eq:BS2}\\
        rt'+r't&=0.  \label{eq:BS3}
    \end{align} \label{eq:BS}
    \end{subequations}

    The coupling of the waveplate is
\begin{equation}
    M_{\text{waveplate}}=
    \begin{pmatrix}
    \cos^2\delta+ e^{-i\theta} \sin^2\delta & (e^{-i\theta}-1)\sin\delta \cos\delta \\
    (e^{-i\theta}-1)\sin\delta \cos\delta & e^{-i\theta}\cos^2\delta+ \sin^2\delta
    \end{pmatrix}. 
\end{equation}

    This satisfies the relations set in Eq.~\eqref{eq:BS}, and therefore could be considered similar to the beam-splitter.

\subsection{Covariance matrix: rotation and losses}\label{Ap:covariance}

        The covariance matrix of the ENMO is given by

             \[ \sigma_{\text{ENMO}} = \]
             \begin{equation}
        \begin{bmatrix}
        \frac{1}{2} & -G_\a \\ \\
        -G_\a & (\frac{1}{2} 
            +\frac{G^2_\a}{2 } \left\lvert\chi_\a\right\rvert^2+\frac{G_\a \kappa_\a |\chi_\a|^2}{2 }\left(\frac{\omega^2 + \kappa_\a^2/4}{\Delta_\a^2}+1\right))
        \end{bmatrix}
         \end{equation}
        
        where the (2,2) term denotes the phase quadrature variance. This term is used for the fits after the rotation transformation and including losses.
        
        To rotate the matrix to get the variance at a particular detection angle $\psi$, and add losses corresponding to the efficiency $\eta$, one can use the following transformation
        
        \begin{align}
           \sigma_{\text{rot}} = R(\psi) \, \sigma_{\text{ENMO}} \, R(\psi)^\top \\
        \sigma_{\text{final}} = \eta \, \sigma_{\text{rot}} + (1 - \eta) \cdot \frac{1}{2} \mathbf{I} 
            \end{align}
        
        where 
           \begin{equation}
               R(\psi) = 
        \begin{bmatrix}
        \cos\psi & \sin\psi \\
        -\sin\psi & \cos\psi
        \end{bmatrix}
           \end{equation}

    \section{Experiemntal setup}

\subsection{Locking loops and techniques} \label{locks}

\begin{figure}[]
            \centering
            \includegraphics[width=\linewidth]{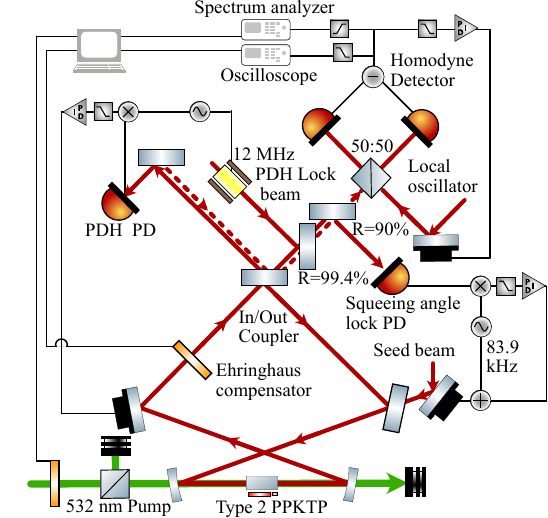}
            \caption{The control loops and auxiliary beams required to operate the ENMO. }
            \label{fig: LockingSetup}
        \end{figure}

    Figure~\ref{fig: LockingSetup} shows the control loops and auxiliary beams required to operate the ENMO. For the p-pol mode, the cavity is kept on resonance using a counter-propagating weak field, which is detected on the cavity's reflection. 12 MHz phase modulation sidebands are introduced using an EOM and then demodulated to get the Pound-Drever-Hall error signal \cite{blackIntroductionPoundDrever2001}. An HR mirror attached to a piezo then controls the cavity length. 
    
    The cavity is detuned for the s-pol mode by changing the birefringence caused by the nonlinear PPKTP crystal. The crystal's fast and slow axes are aligned to the cavity axes as shown in Appendix~\ref {crystal0}. The crystal is wedged at both end surfaces to \qty{1}{\degree}. Moving the crystal perpendicular to the optical axis is used to change the crystal length in the beam and therefore the effective retardation between the two polarization modes. Further fine adjustment is done by changing the crystal edge temperature and thereby changing the effective retardation. The bulk of the crystal is controlled at the temperature for the best down-conversion efficiency and is not affected by the edge temperature changes.

    To lock the squeezing angle, a weak p-pol seed beam is sent to the cavity and co-propagates with the squeezed signal. A pick-off of the signal is sent to a photodiode to see the amplification and de-amplification of the seed beam. Dithering the seed beam creates a phase modulation between the seed and the pump beam at $2\pi \times$\qty{83.9}{kHz}. Demodulating the photodiode signal, helps obtain the error signal to lock at the amplified or de-amplified peaks.

    The homodyne detection angle is locked using the error signal from the seed and the local oscillator interference. The seed dithering mentioned above is used for the same. 

    The precise control of the pump beam power and therefore the $g_\DC$ is changed using a computer-controlled motorized waveplate rotation and a polarization beam-splitter. The precise control of the Ehringhaus compensator rotation and therefore the $g_\BS$ is similarly done with another motorized rotation mount.

      \begin{table}[h]
            \centering
                           \begin{tabular}{ l r  }
            \\
            \toprule
            Efficiency   &  Value \\ \midrule     
            Propagation efficiency  & \qty{91}{\percent} $\pm$ \qty{4}{\percent} \\ 
            Homodyne balancing &  \qty{99.9}{\percent}  $\pm$ \qty{0.1}{\percent}  \\  
           Homodyne contrast  & \qty{90}{\percent} $\pm$ \qty{4}{\percent} \\ 
           Quantum efficiency & \qty{97}{\percent}$\pm$ \qty{2}{\percent}\\
            Escape efficiency & \qty{68.4}{\percent}$\pm$ \qty{0.5}{\percent}\\ \midrule  
            Total efficiency & \qty{54}{\percent}$\pm$ \qty{4}{\percent}\\ \bottomrule
            Squeezing & \qty{-2.6}{dB}$\pm$ \qty{0.1}{dB}\\  
            Anti-squeezing & \qty{6.0}{dB}$\pm$ \qty{0.1}{dB}\\  \midrule  
             Measured efficiency & \qty{53.}{\percent}$\pm$ \qty{2}{\percent}
            \\ \bottomrule
            \end{tabular}
            \caption{Shows the efficiency values for different loss channels and the corresponding calculated total efficiency. The measured efficiency when the system is operated as a single-mode squeezer is consistent with the total efficiency.}
            \label{tab: loss budget}
            \end{table}

    \subsection{Loss budget}

        In order to budget the losses in the cavity and minimize the unaccounted losses, the system is operated as a simple single-mode squeezer. This is by setting the waveplate to the highest coupling state, where the p-pol mode is entirely converted to s-pol and vice versa. This is analogous to the power-coupled cavity scheme with the coupling mirror of \qty{100}{\percent} transmission. This essentially doubles the cavity length, halving the FSR. The Tab.~\ref{tab: loss budget} shows the efficiencies corresponding to different loss channels when the system is operated as a simple single-mode squeezer. The measured efficiency from the squeezing and anti-squeezing measurements are consistent with the total efficiencies

\subsection{Techniques to avoid the down-conversion crystal causing beam-splitter-like interaction.}\label{crystal0}

    The nonlinear PPKTP crystal used for the two-mode squeezing process in the ENMO cavity is birefringent. A birefringent crystal has a fast axis and a slow axis orthogonal to each other. When these axes are tilted with respect to the cavity polarization mode basis, the birefringent crystal acts as a waveplate. However, since the beamsplitting process inside the cavity is of importance and has to be carefully selected by the Ehringhaus compensator, any residual beam-splitting process by the nonlinear crystal is undesired. Therefore, techniques are to be developed to eliminate the residual beam-splitting process from this crystal. 

    The beam-splitting coupling strength $g_\BS = \sin(2\delta)\,  \theta\, \textit{FSR}$ could be set to zero by either setting $\delta=0$ or $\theta=0$. Because we use $\theta$ to control the phase difference between the two cavity polarization modes, we opt to set $\delta=0$. This is done by rolling the crystal in the axis of the laser beam propagation, similar to rotating a waveplate. However, setting $g_\BS$ precisely to zero requires accurate ways to measure or verify it. 

    There are two ways to monitor while reducing the $g_\BS$. A coarse method is to minimize the normal mode splitting frequency while the cavity length is being ramped \cite{steinmeyerSubsystemsAllopticalCoherent2019}. The normal mode splitting frequency is proportional to $g_\BS$. The second, much more sensitive method is to pump the down-conversion crystal and verify that the detected quantum state is a thermal state. Pumping the down-conversion crystal creates correlated sidebands between s-pol and p-pol modes. However, in the case of $g_\BS = 0$, the s-pol sideband mode entirely decays into losses and does not enter the detected field. Therefore, the detected field will be a thermal state or just one part of the EPR pair. Detecting the quantum state at different detection angles and minimizing the fluctuation of the variance over different detection angles minimizes the $g_\BS$.

\end{appendix}

\input{bibliography.bbl}
\end{document}

%% file: bibliography.bbl
%